\documentclass[]{spie}  
\usepackage[]{graphicx}
\usepackage[lofdepth,lotdepth]{subfig}

\title{Characterizing the red optical sky background fluctuations from narrow-band imaging} 

\author{M. Puech\supit{a}, H. Flores\supit{a}, Y.B. Yang\supit{a,b}, M. Rodrigues\supit{c,a,d}, T. Gon\c calves\supit{e}, F. Hammer\supit{a}, and K. Disseau\supit{a}
\skiplinehalf
\supit{a}GEPI, Observatoire de Paris, CNRS-UMR8111, Univ. Paris-Diderot, 5 place Janssen, 92195 Meudon, France\\
\supit{b}National Astronomical Observatories, Chinese Academy of Sciences, 20A Datun Road, Chaoyang District, Beijing 100012, China\\
\supit{c}European Southern Observatory, Alonso de Cordova 3107 - Casilla 19001 - Vitacura -Santiago, Chile\\
\supit{d}CENTRA, Instituto Superior Tecnico, Av. Rovisco Pais 1049-001 Lisboa , Portugal\\
\supit{e}Observatorio do Valongo, UFRJ, Ladeira Pedro Antonio, 43 - Saude, Rio de Janeiro-RJ, CEP 20080-090, Brazil\\
}

\authorinfo{Send correspondence to M. Puech: mathieu.puech@obspm.fr}

\begin{document} 
  \maketitle 

\begin{abstract}
  The detection and characterization of the physical properties of
  very distant galaxies will be one the prominent science case of all
  future Extremely Large Telescopes, including the 39m E-ELT.
  Multi-Object Spectroscopic instruments are potentially very
  important tools for studying these objects, and in particular
  fiber-based concepts. However, detecting and studying such faint and
  distant sources will require subtraction of the sky background
  signal (i.e., between OH airglow lines) with an accuracy of
  $\sim$1\%. This requires a precise and accurate knowledge of the sky
  background temporal and spatial fluctuations. Using FORS2
  narrow-band filter imaging data, we are currently investigating what
  are the fluctuations of the sky background at $\sim$9000A. We
  present preliminary results of sky background fluctuations from this
  study over spatial scales reaching $\sim$4 arcmin, as well as first
  glimpses into the temporal variations of such fluctuations over
  timescales of the order of the hour. This study (and other
  complementary on-going studies) will be essential in designing the
  next-generation fiber-fed instruments for the E-ELT.
\end{abstract}



\section{INTRODUCTION}
\label{sec:intro}  
One of the prominent science cases for the future ESO 39 meter European
Extremely Large Telescope (E-ELT)\cite{gilmozzi11} is the detection
and characterization of the first galaxies at very high redshifts
(z$\leq$6)\cite{evans12}. Spectroscopic observations of such small
(half-light radii R$_{half}\sim$0.2 arcsec) and faint
(J$_{AB}\sim$26-27) objects in the Near-Infrared (Y to Ks bands), will
require accurate and precise background subtraction techniques at a
level of one percent\cite{navarro10}. This puts significant constraints
on the instrument design concept and operations.


Fiber-fed instruments are often believed to suffer from a disadvantage
in terms of sky subtraction accuracy compared to slits or image slicer
IFUs. The two main drawbacks of fibers are thought to be their global
throughput, which is limited by Focal Ratio Degradation (FRD), and the
necessity of measuring the background signal several arcsec away of
the scientific target because of the finite mechanical extension of
the fiber buttons on the focal plane. While fiber-fed instrument can
now offer global throughputs similar to those of slit
spectrographs\cite{navarro10}, it remains to be demonstrated that the
background signal can be subtracted with an accuracy of at least one
percent.

In this paper, we used narrow-band archive imaging data obtained with
ESO-VLT/FORS2 to characterize the sky background spatial and temporal
fluctuations. This should bring useful constraints on the maximal
distance between the scientific target and an associated fiber
dedicated to measuring the sky background signal. In a companion
paper, we also explored spectroscopic long-slit observations from
ESO-VLT/FORS2\cite{yang12}. We also conducted on-sky tests of
different sky background measurement and correction techniques using
the multi-fiber optical spectrograph
ESO-VLT/FLAMES-GIRAFFE\cite{rodrigues12}. The reader is directly
referred to these companions papers for further information on these
on-going parallel studies.

\section{Objectives and methodology of the study}
Very distant galaxies have their emission lines redshifted in the NIR.
They are selected in order to target these lines between the very
bright OH airglow sky lines. The detection success rate of very high-z
faint sources is therefore driven by the accuracy on the measurement
and subtraction of the sky background continuum between the OH sky
lines. First tests on the sky subtraction strategy were conducted and
are described elsewhere\cite{rodrigues12}. In the present study, we
focus on the measurement issue and we characterize the spatial and
temporal sky background fluctuations in the continuum. Because longer
wavelengths are more affected by thermal emission from the telescope
and instrument, we decided to first investigate the red part of the
optical domain ($\lambda\sim$900 nm). Similar data at longer
wavelengths will be explored in future studies.

\begin{figure}[h]
\begin{center}
\includegraphics[width=10cm]{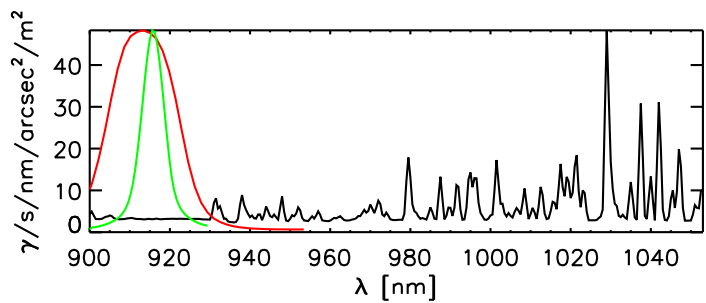}
\end{center}
\caption{\label{FigFilters}Narrow-band filters used in the archive
  data. A NIR sky background model for Mauna Kea is represented in
  black (airmass=1.5, middle of night). The red curve represents the
  z\_SPECIAL transmission curve, while the green one represents the
  FILT\_917\_6 transmission curve (in arbitrary units).}
\end{figure}

We looked into the ESO raw data archive for programmes using
narrow-band images at red optical wavelengths, and we find a number of
ESO-FORS2 programmes suitable for our project. We further narrowed
down the search to programmes that offered a large number of exposures
with at least $\sim$1 hr of continuous observations per night over
several nights, and with large enough individual integration times
(i.e., at least several minutes) to avoid the read-noise limited
regime in the background. We ended up with two programmes particularly
well-suited for our purposes, whose main characteristics are
summarized in Tab. \ref{TabProg}. We checked that no strong OH sky
lines were in the spectral bandwidth of the two respective narrow-band
filters, as shown in Fig. \ref{FigFilters}.

\begin{table}[h]
  \caption{ESO-VLT/FORS2 programmes used in the present study. \emph{From left to right:} Programme ESO ID, programme PI, narrow-band filter used, central wavelength of the filter in nm, FWHM of the filter in nm, individual detector integration time DIT, and pixel size in arcsec.} 
\label{TabProg}
\begin{center}       
\begin{tabular}{ccccccc}\hline
ID & PI & Filter & $\lambda _c$ & $\Delta \lambda$ & DIT & Pixel\\\hline
68.A-0182 & A. Cimatti & z\_SPECIAL & 915 & 20 & 380 & 0.25\\
70.A-0591 & E. Daddi & FILT\_917\_6 & 917 & 6 & 360 & 0.25\\\hline
\end{tabular}
\end{center}
\end{table} 

\section{DATA REDUCTION}

We downloaded raw data from the ESO archive for the two programmes
listed in Tab. \ref{TabProg}. For homogeneity, we did not consider the
first night (2001-24-11) of Prog. 68.A-0182 which was conducted before
the change of CCD in FORS2 that occurred in March
2002\cite{moehler10}. We checked the availability of calibration
frames (bias and twilight flat fields [hereafter FF]) in the archive
for each night of both programmes. This led us to discard one night of
Prog. 70.A-0591, because we were not able to retrieve daily bias
frames for the remaining night, and one night (2002-10-07) of Prog.
68.A-0182, for which we were able to retrieve associated twilight FF
frames.

\subsection{Basic steps of data reduction}
The basic steps of the standard reduction process were conducted using
GASGANO\cite{izzo04} and the associated ESO recipes. We first
constructed master bias frames using the five to ten available daily
raw bias frames in the archive. Master FF frames were also constructed
using the associated four to five available daily twilight FF frames
in the archive. Each individual raw frame was then reduced using
standard procedures. We did not apply any dark correction since this
is usually useless with FORS2 data\cite{obrien07} but we checked that
no particular spatial structure was imprinted in the dark frames
available in the archive over the programme period. Reduced individual
science frames were finally cut to remove the non-exposed part of the
image and prevent possible edge effects during further steps. After
visual inspection of the raw and reduced science frames, we
systematically discarded data associated with the second (slave) CCD
chip, since they were found to be more affected by cosmetic issues
(e.g., bad pixels and/or columns). Regarding Prog. 68.A-0182, we
further discarded data associated with the fourth night of the
programme (2002-11-09), since they were affected by many satellite
glows. This lets us with 35 frames spread over 4 nights (2002-10-08,
2002-11-11, 2003-02-03, and 2003-02-23). The individual frames sample
periods of $\sim$1 to 2 hr per night with sizes of approximatively
7$\times$4 arcmin$^2$. Regarding Prog. 70.A-0591, the final dataset is
made of 63 raw frames, which sample periods of $\sim$1 to 2 hr per
night spread over four nights (2003-02-05, 2003-06-02, 2203-02-08, and
2003-10-02). Unfortunately, we were not able to retrieve associated
daily twilight FF frames in the archive for this programme, so we used
those taken on the 2003-09-02, which were used to construct a master
FF frame. Since the four nights of observation considered were almost
consecutive, we reduced all the individual frames using this master
FF.

\subsection{Background models and superflat-fielding}
We used SWARP\cite{bertin10} to combine all the dithered individual
reduced science frames into master field mosaic images.
SEXTRACTOR\cite{bertin96} was then used to detect all the objects in
these images. We used very low detection thresholds (0.3 to
0.7$\sigma$) combined to a Gaussian filter with large kernel (3 pixels
FWHM) to ensure that the residual light from extended objects
(galaxies) in the field is minimized. An additional morphological
filtering was performed to enlarge the masks and limit the pollution
by diffuse light at the edge of the objects in the field. Masks were
projected back onto the individual reduced science frames, which were
then normalized by the median value of the background (i.e., pixels
not masked during the previous step). All normalized images were
combined using a sigma-clipped median in the detector pixel frame
(i.e., pixel-to-pixel) to produce a background model per night, in
units of the median background level.

\begin{figure}[h]
\begin{center}
\includegraphics[width=17cm]{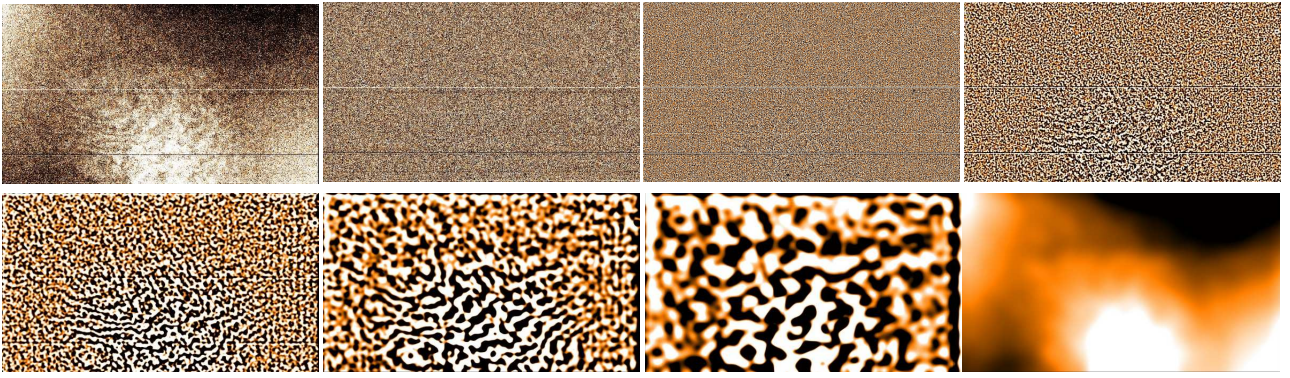}
\end{center}
\caption{\label{FigBack}Scale decomposition of the background model
  for Prog. 70.A-0591. Here we show the background model constructed
  using all the reduced science frames, but daily background models
  were used to superflat the data. \emph{From left to right and top to
    bottom:} background model, successive decomposition at scales
  $2^k$ pixels with k=0-5, and residual large-scale variations
  (subtraction of the k=0 to 5 scales from the background model).}
\end{figure}

\begin{figure}[h]
\begin{center}
\includegraphics[width=17cm]{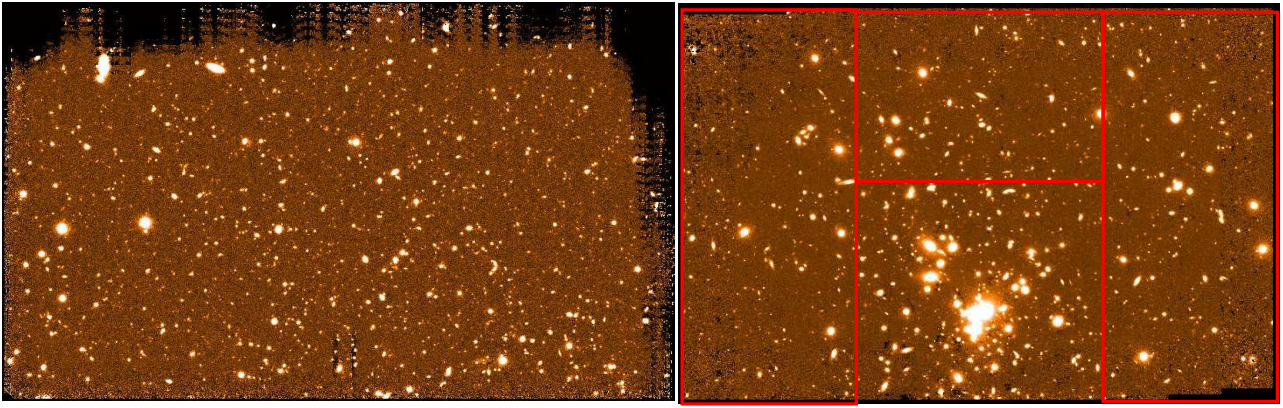}
\end{center}
\caption{\label{FigMasterFields}Master Fields (chip 1) of the two
  FORS2 programmes considered in this study. The master fields were
  constructed using SWARP from all the dithered super-reduced science
  frames. \emph{Left}: programme 68.A-0182, for which all the field
  was analyzed. \emph{Right:} programme 70.A-0591, for which only the
  three regions on the right, left, and top of the image (see red
  rectangles) were analyzed separately in order to avoid the densest
  region at the bottom center of the image (see text). Black pixels
  are pixels for which the superflats are incomplete due to the
  limited number of dithered science frames.}
\end{figure}

Each background model was analyzed using a wavelet
decomposition\cite{starck06} and visually inspected for possible
additive features such as scattered light or fringing. Fringing at
$\lambda \sim$900 nm is expected at scales of $\sim$40 pixels for the
closest fringes\cite{walsh08}, and with peak-to-valley intensities of
a few percent depending on exposure time\cite{walsh08}. Using a scale
decomposition of a background model, one could in principle separate
additive and multiplicative residual light as follows (see Fig.
\ref{FigBack}). The highest scales (k=0-1, which contain variations
over $2^k$=1-2 pixels) allows us to characterize the noise and
cosmetic properties (e.g., bad pixels and/or columns, noise patterns).
The fringe pattern starts appearing at scales k=2 (variations over 4
pixels) and remains evident up to scale k=6 (i.e., variations over 64
pixels). The distance between the fringes is found to be 15-20 arcsec
with peak-to-valley amplitudes of $\sim$0.1-0.3\% consistent with
expectations at these wavelengths\cite{walsh08}. For both programmes,
the fringe pattern is found to be significant only in the central
bottom part of the CCD, where the density of galaxies in the field is
the highest (see Fig. \ref{FigMasterFields}). Fringing is due to
multiple reflections of quasi-monochromatic light between the surfaces
of the CCD\cite{howell12}. Because it is associated to reflections and
not to transmission effects in the instrument, it should in principle
be corrected additively\cite{gull92}. One can therefore sum up the
scale in which the variations are dominated by the fringe pattern (k=2
to 5), and subtract the result from the reduced frames. The residual
large-scale variation can then be used as a second order
multiplicative flat-field correction.

Background models revealed significant daily variations, so we did not
consider long-term models but constructed daily background models.
Given the limited number of calibration frames, each step of the data
reduction is likely to add significant noise to the individual science
frames\cite{gull92,newberry91}. Since the fringe pattern is found to
be relatively weak, we chose not to apply an additive correction to
save one reduction step and preserve the signal-to-noise ratio in the
background. Instead, we treated the daily background model as a
superflat used to apply a second-order flat-field multiplicative
correction for both programmes. In doing so, one expects that sky
background features should average in the detector frame, provided
that their size is smaller than the image size. Such structures should
therefore be preserved during the super-flatfielding correction.
Further analysis of the background spatial fluctuations is therefore
limited to scales that are typically smaller than half the image
diagonal, according to the Nyquist-Shannon rule. For Prog. 68.A-0182
this translates into a maximal scale of $\sim$240 arcsec. For Prog.
70.A-0591 we used the master field image produced using SWARP to
define three rectangular regions around the central bottom region in
the images where the fringe pattern is the strongest, as illustrated
in the left panel of Fig. \ref{FigMasterFields}. We found a
significant amount of diffuse light in this region which was indeed
difficult to mask entirely using SEXTRACTOR because of the very high
density of galaxies. Masking this region therefore prevents us to
introduce large amounts of diffuse light in subsequent steps. Since we
analyzed only three subregions, this programme allows us to
characterize the spatial fluctuation of the background only up to
scales $\sim$140 arcsec.

The master field images resulting from the combination of all the
super-reduced science frames are shown in Fig. \ref{FigMasterFields}.
This figure illustrates that the limited number of dithered frames per
night (typically 8-10 frames) did not allow us to construct complete
superflat frames per night, i.e., superflat frames still contained
blank pixels. These blanks pixels result from regions that were masked
on all the individual frames because of an extended bright object
lying over these pixels in all the images.

\subsection{Background continuum maps}
\begin{figure}[h]
\begin{center}
\includegraphics[width=17cm]{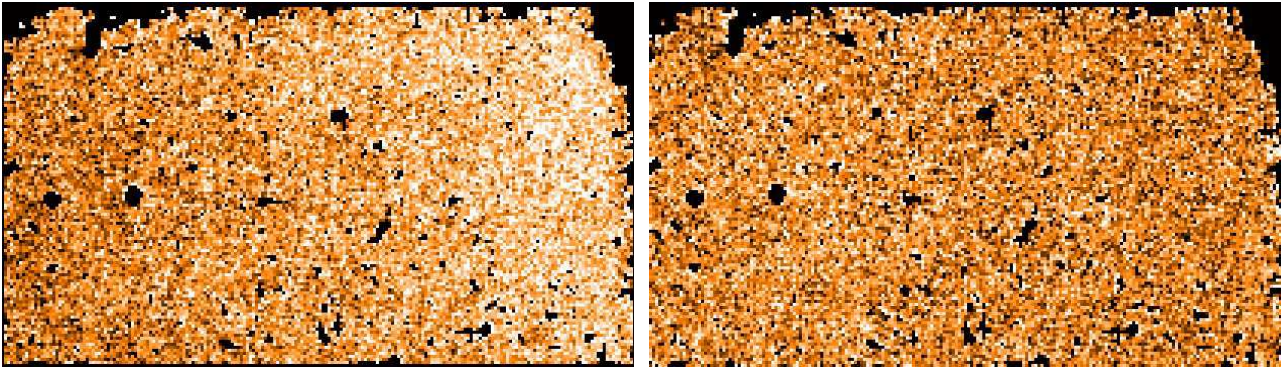}
\end{center}
\caption{\label{FigSkyMaps}Two examples of background maps obtained
  using the programme 68.A-0182 (2 arcsec/pixel). Values are typically
  ranging between $\sim$ 0.995 and 1.005.}
\end{figure}

To reduce the impact of noise on the background fluctuations, we
``rebinned'' the super-reduced science frames as follows. Using the master
field images, a 2$\times$2 arcsec$^2$ grid sampling the total observed
field-of-view was constructed for each programme. This grid was
projected back onto each super-reduced science image. Within each
2$\times$2 arcsec$^2$ box of the grid, a sigma-clipped background
median value was calculated to construct a background map for each
individual science image. We therefore ended up with a set of
background maps as a function of time with sampling of 2 arcsec. We
show of few examples of such background maps in Fig. \ref{FigSkyMaps}.
Because we are interested only in relative fluctuations, we did not
try to flux-calibrate the data. Since each super-reduced image was
normalized by the median background value, the background maps allow us
to characterize the fluctuations of the background signal
\emph{relative} to the median background value. We refer to other
studies for an absolute characterization of the sky background
continuum\cite{cuby00}.

We inspected the noise properties of the background maps before (0.25
arcsec/pix) and after rebinning (2 arcsec/pix). The median r.m.s.
value before rebinning is found to be 3.9 and 16.4\% of the median
value for Prog. 68.A-0182 and 70.A-0591 respectively. This difference
should be in part due to the narrower filter and slightly shorter
integration time used in Prog. 70.A-0591 (see Tab. \ref{TabProg}),
which are expected to result in a larger photon noise. The rebinning
process should reduce the photon noise by a factor of at least
$\sqrt{64}$, which should lead to r.m.s. values $\sim$0.5 and 2\%
respectively. We measured the r.m.s. fluctuations in the rebinned
background maps, which are found to be 0.3 and 0.8\% respectively. The
sigma-clipping used to generate background maps during the rebinning
process therefore mitigates the photon noise efficiently. This
suggests that background maps are not dominated by photon noise but by
fluctuations at the smallest scale (see Fig. \ref{FigSkyMaps}),
although it cannot be excluded that residual photon noise still
contribute to the r.m.s. fluctuations of the normalized background
maps (see below).

Finally, we inspected visually all the super-reduced images as well as
the daily superflat images and note that no concentric features were
present. This suggests that the produced background maps are not
dominated by scattered light or well-known sky concentration effects
(i.e., light suffering multiple reflections between optical surfaces
before reaching the detector), although it remains difficult to assess
whether or not the background maps contain pure \emph{sky} background
signal or additional residual light from the instrument. However, analyzing
such background maps may provide useful upper limits on the sky
background fluctuations.

\section{Autocorrelation analysis of the background continuum maps}
\begin{figure}[h]
\begin{center}
\includegraphics[width=4.1cm]{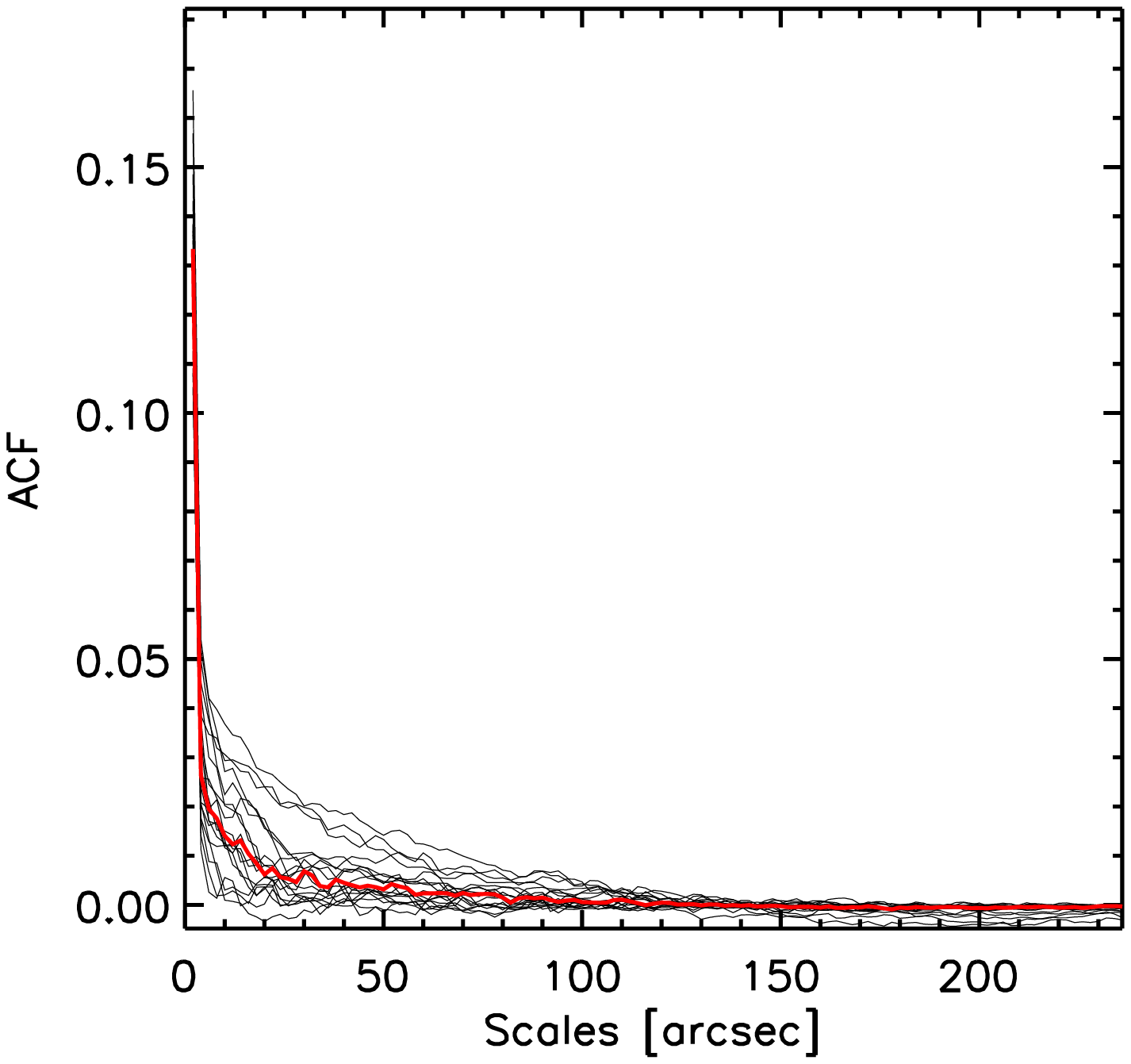}
\includegraphics[width=4.1cm]{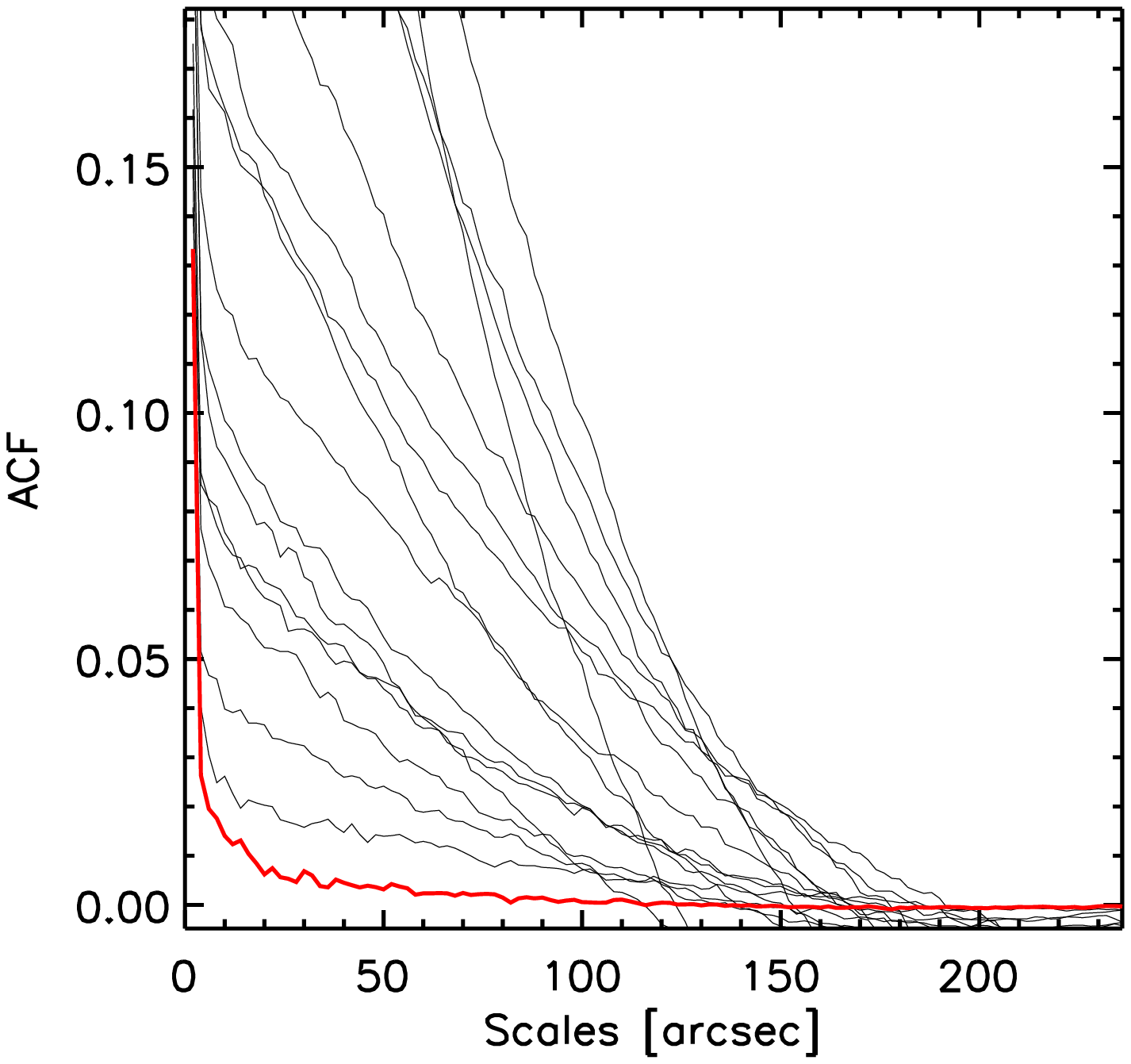}
\includegraphics[width=4.1cm]{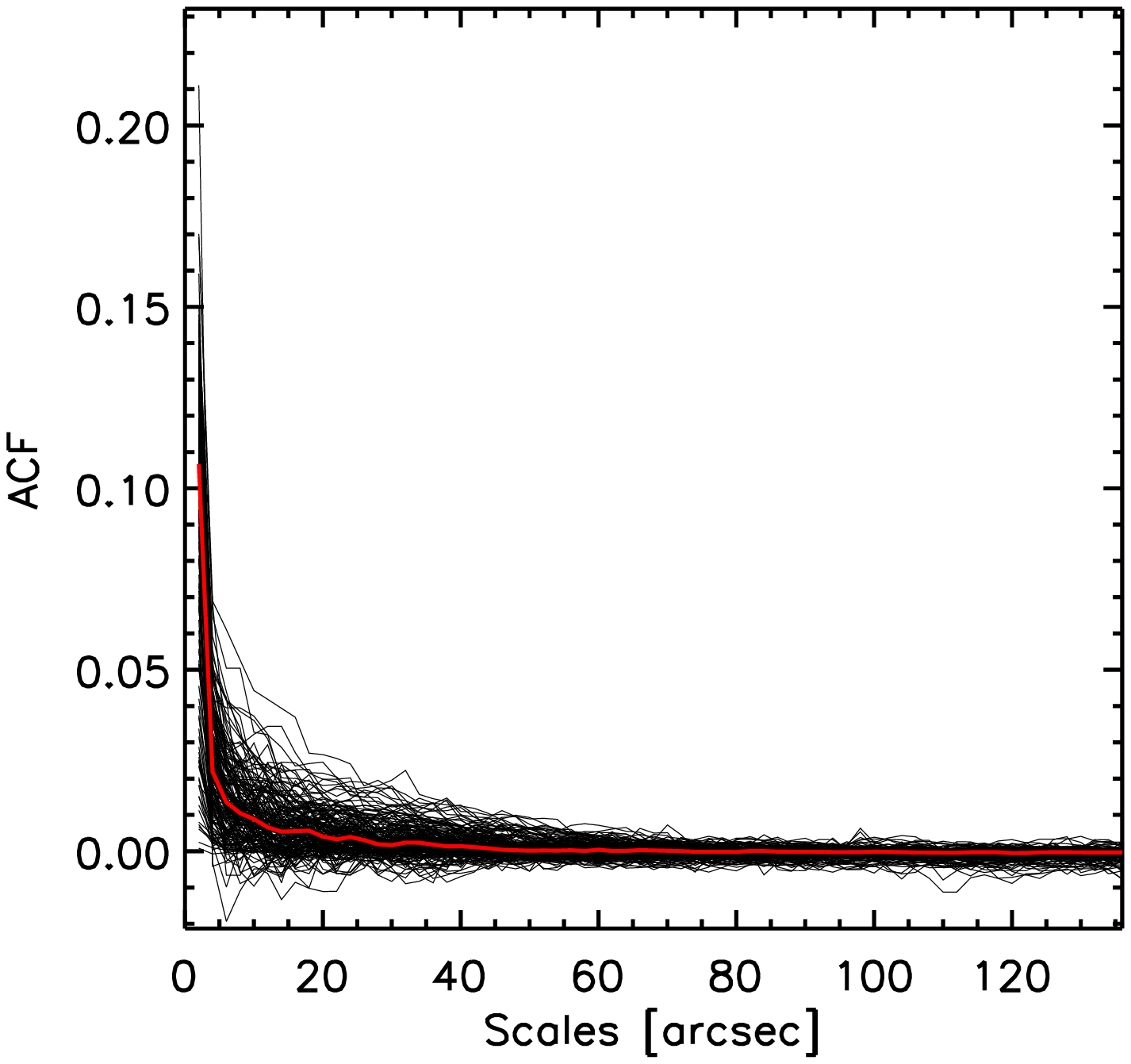}
\includegraphics[width=4.1cm]{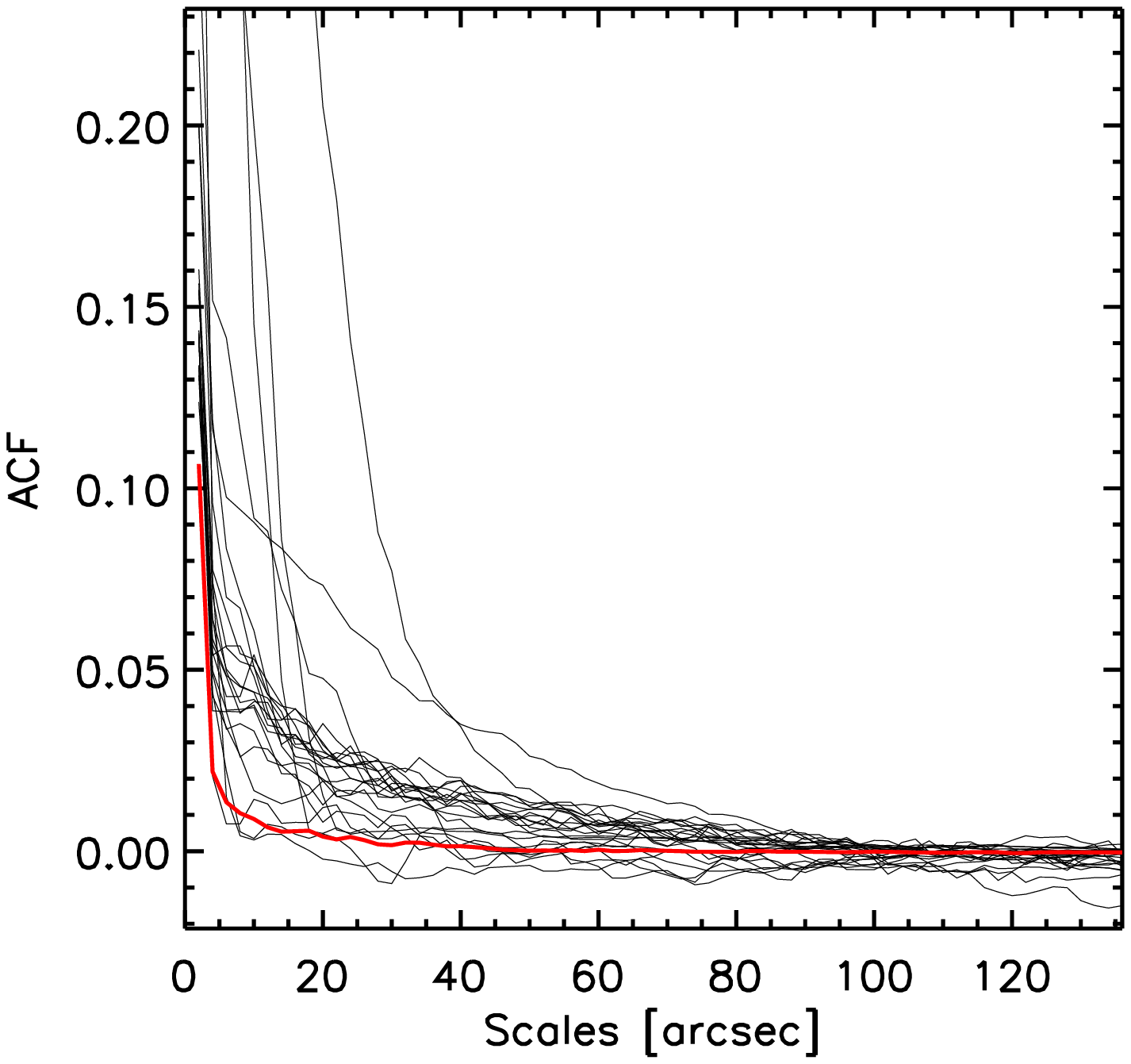}
\end{center}
\caption{\label{FigACFAll}\emph{From left to right:} total 68.A-0182
  ACF and rejected ACFs for Prog. 68.A-0182 and 70.A-0591
  respectively. Black lines in the first and third panels represent
  ACFs derived from individual background maps that were not rejected
  by the consistency criterion (see text), while the red line is the
  corresponding median ACF. Black lines and the second and fourth
  panels show the ACFs rejected by the consistency check criterion,
  while the red line is the median ACF from the first and third
  panels. The value of the ACFs at origin was not plotted for
  visualization ease (ACF(0)=1).}
\end{figure}

The autocorrelation function (ACF) of each background map was derived
up to the limiting spatial scale (see above). In the following, we
assume that the relative background signal is stationary over the
considered temporal and spatial scales. We checked that the mean and
r.m.s. relative background values were approximatively stable over the
field-of-view, which suggests that the relative background signal is
roughly stationary at least up to the second order, i.e., wide-sense
stationary (WSS). We therefore assumed that the ACF depends on spatial
shift only. Because the resulting ACFs are quite noisy, we considered
daily median-averaged ACFs rather than individual ACFs. Since the
individual exposure time and sampled period per night were more
favorable for the programme 70.A-0591, we tentatively tried to
construct hourly median-averaged ACFs for this programme only. We
finally constructed two median-averaged ACFs considering all the
frames of each programme, in order to check for possible systematics
between the two different filters (see Fig. \ref{FigACFAll}). In the
following, we will refer to these different temporally average ACFs as
daily, hourly, and total ACFs respectively. We now detail how
median-averaged ACFs were constructed.

To mitigate the impact of noise, we used an iterative procedure to
construct median-averaged ACFs. For each spatial scale, we derived the
median value of the median-subtracted, r.m.s.-normalized set of
relative background maps considered. The ACF of each individual
background map $ACF$ was then compared to the median ACF $Median
_{ACF}$ using the following consistency function $C$:
$$
C(scales)=\sum _{scales} \frac{(Median _{ACF}-ACF)^2}{Median _{ACF}.N_{scales}}
$$ The sigma-clipped median value $C$ over all scales of $C(scales)$
was then derived. Every individual $ACF$ having $C\geq$ 5 was
discarded. The process was iterated until no further rejection
occurred. Tab. \ref{TabFilt} lists the fraction of background maps
discarded from the calculation of the median ACF per set of data
considered. For Prog. 70.A-0591, the background maps corresponding to
the three different regions extracted from the images were analyzed as
distinct maps. Note that we did not average the data directly, which
would have smooth out spatial features with short timescales. Instead,
we averaged the ACF resulting of each background map, we should
preserve statistically the spatial features of each background map.
Table \ref{TabFilt} shows that there is a non negligible variability
of the ACF at all sampled timescales (0.5 to 3.5 hr). Given the
relatively low signal-to-noise ratio of the relative background
signal, it is difficult to test whether this variability can be
attributed to pure background fluctuations or are due to noise
fluctuations. However, in case of sky measurements using the offset
technique, this suggests that the background signal should be sampled
at timescales smaller than 0.5 hr, at least in the red optical
($\lambda \sim$900 nm).

\begin{table}[h]
  \caption{Rejection rate of the iterative procedure used to construct the median ACFs for each dataset. \emph{From left to right:} Dataset (Total = all background maps for a given programme, YYYY-MM-DD = all maps for a given night/programme), YYYY-MM-DDTHH = all maps within HH to HH + 1 hr of a given night/programme), rejection rate in \%, remaining number of background maps used to construct the ACF, and approximative timescale sampled by the individual maps in hr.} 
\label{TabFilt}
\begin{center}       
\begin{tabular}{|cccc|cccc|}\hline
{\bf 70.A-0591} & & & & {\bf 68.A-0182} & & & \\\hline
DataSet & Rejection rate & N$_{ACF}$ & Timescale & DataSet & Rejection rate & N$_{ACF}$ & Timescale\\\hline
Total & 46\% & 166 &-- & Total & 12\% & 19&-- \\\hline
2003-02-05 & 11\% & 24 & 1 & 2002-10-08 & 9\% & 10& 1\\
2003-02-06 & 22\% & 42 & 2 & 2002-11-11 & 0\% & 8& 1 \\
2003-02-08 & 16\% & 63 & 3.5 & 2003-02-03 & 0\% & 8& 1\\
2003-02-10 & 4\% & 26 & 1 & 2003-02-23 & 0\% & 8& 1\\\hline
2003-02-05T07 & 0\% & 15 & 0.5 & & &&\\
2003-02-05T08 & 0\% & 12 & 0.5 & & &&\\
2003-02-06T07 & 29\% & 17 & 0.5 & & &&\\
2003-02-06T08 & 8\% & 22 & 0.5 & & &&\\
2003-02-06T09 & 0\% & 6 & 0.5 & & &&\\
2003-02-08T05 & 8\% & 11 & 0.5 & & &&\\
2003-02-08T06 & 11\% & 24 & 0.5 & & &&\\
2003-02-08T07 & 7\% & 14 & 0.5 & & &&\\
2003-02-08T08 & 10\% & 19 & 0.5 & & &&\\
2003-02-10T07 & 0\% & 18 & 0.5 & & &&\\
2003-02-10T08 & 0\% & 9 & 0.5 & & &&\\\hline
\end{tabular}
\end{center}
\end{table}

\begin{figure}[h]
\begin{center}
\includegraphics[width=5.5cm]{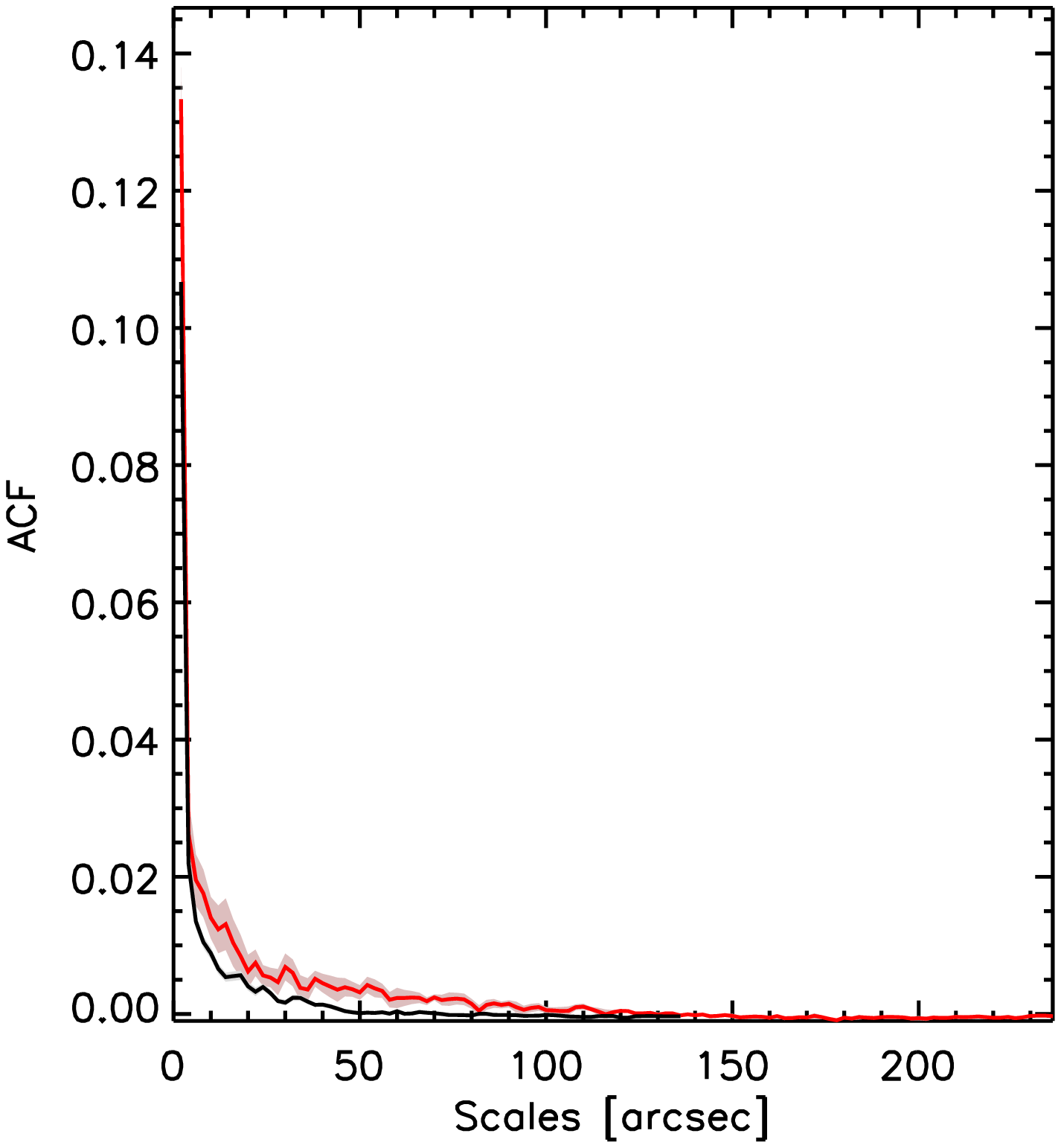}
\includegraphics[width=5.5cm]{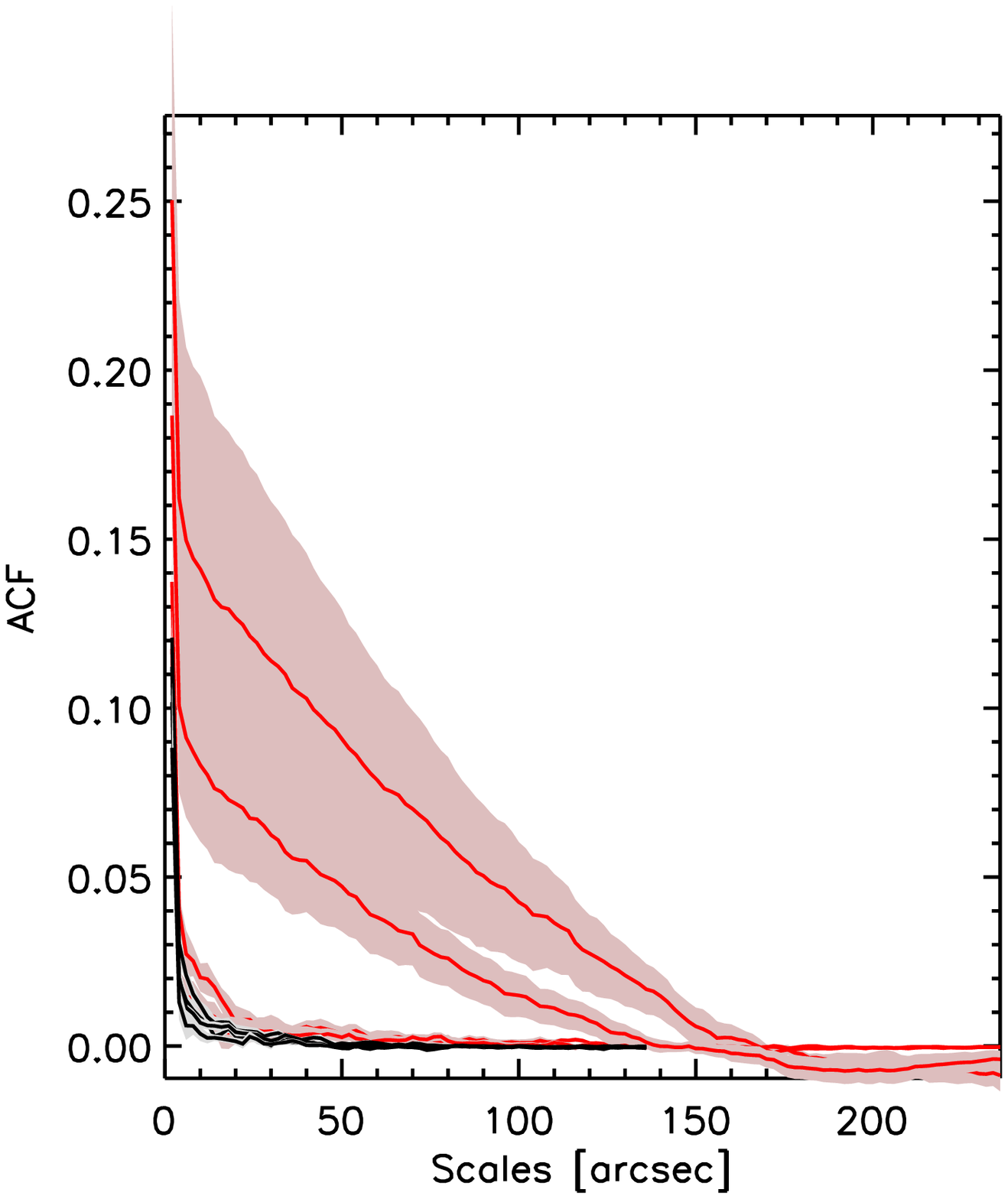}
\includegraphics[width=5.5cm]{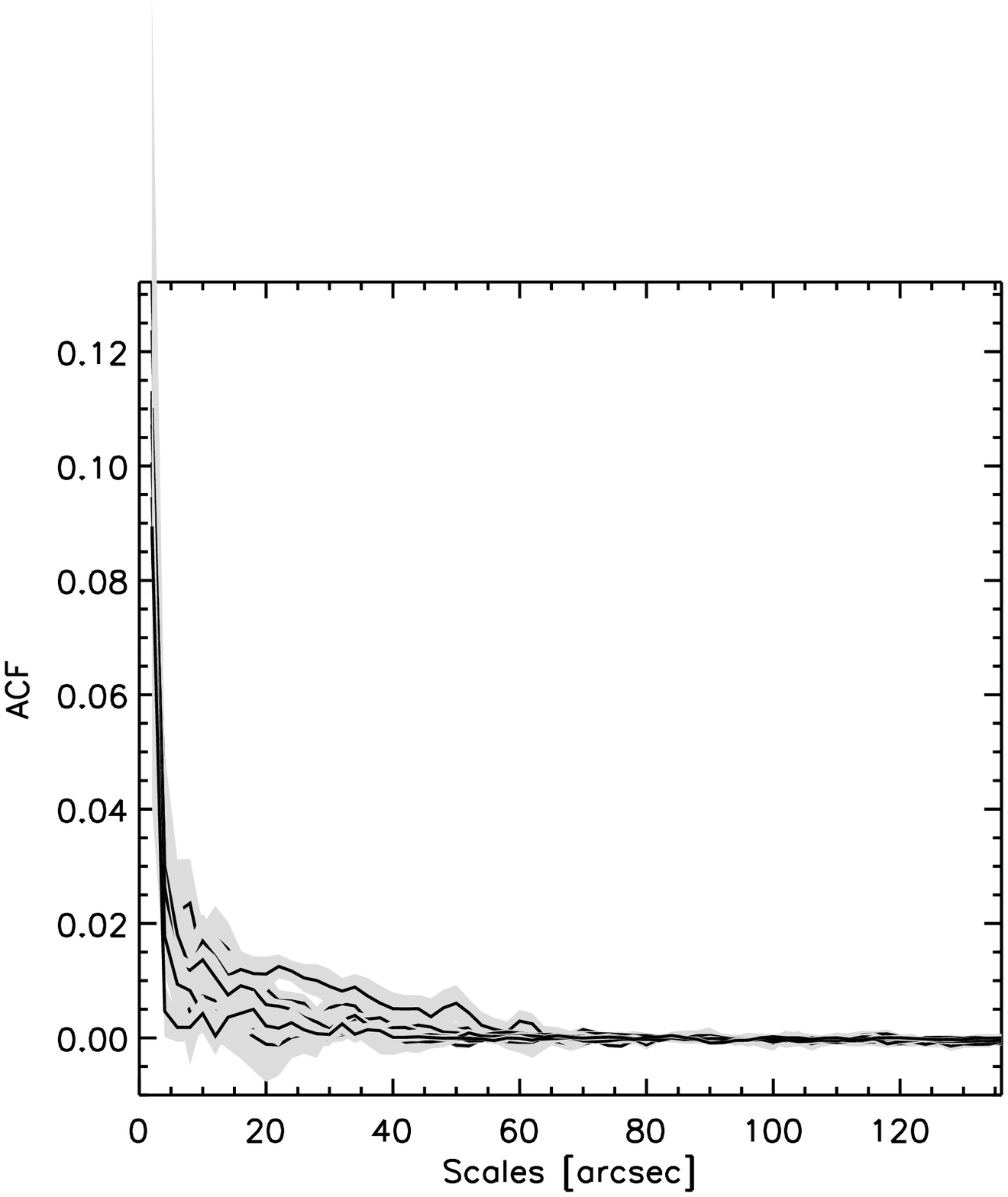}
\end{center}
\caption{\label{FigACFComp}Comparison between the total (left panel),
  daily (middle panel), and hourly (right panel) median ACFs. The red
  lines shows ACFs for the 68.A-0182 programme, while the black lines
  show ACFs for the 70.A-0591 programme. Uncertainties on the median
  ACFs (as derived from bootstrap resampling) are shown as gray or
  light red areas.}
\end{figure}

The median total ACFs are shown in Fig. \ref{FigACFAll} (see also Fig.
\ref{FigACFComp}). They show a very similar behavior, which suggests
that the impact of using different filters is limited. Figure
\ref{FigACFAll} also shows the ACFs that were rejected during the
median calculation. All rejected ACFs show a trend to larger
correlations at all scales.

We compare the total, daily, and hourly ACFs in Fig. \ref{FigACFComp}.
Daily ACFs show consistent behavior except for two nights in Prog.
68.A-0182 (2002-10-08 and 2003-03-23) where larger correlations over
all scales are detected. This shows that such a systematic increase in
correlation at all scales can be transitory but also stable over at
least one hour.

\section{Spatial and temporal fluctuations of the background continuum}

\subsection{Sizes and amplitudes of fluctuation scales}
In order to further quantify the fluctuations of the relative
background signal, we fitted the ACFs corresponding to different
datasets as follows. We first generated Gaussian Random Fields (GRFs)
on a 200$\times$200 pixel grid. We combined several GRFs of different
scales (defined as the FWHM of the Gaussian kernel used to generate
the GRF) and amplitudes (in units of the median background signal),
and calculate the resulting ACF. The scales and amplitudes of the
combined GRFs were then fitted to a given median observed ACF using a
Levenberg-Marquardt $\chi ^2$ minimization scheme. We used the
1-$\sigma$ scatter around the median ACF as uncertainties during the
fitting process. We find that at least three GRFs were needed to
provide a satisfying fitting, while more GRFs did not provide any
significant improvement. We also tried to directly fit the ACF shape
using a set of three Gaussian distributions assuming uncorrelated
Gaussian fluctuations, but we found that the result were less accurate
that with fitting GRFs.

\begin{figure}[h]
\begin{center}
\includegraphics[width=3.1cm]{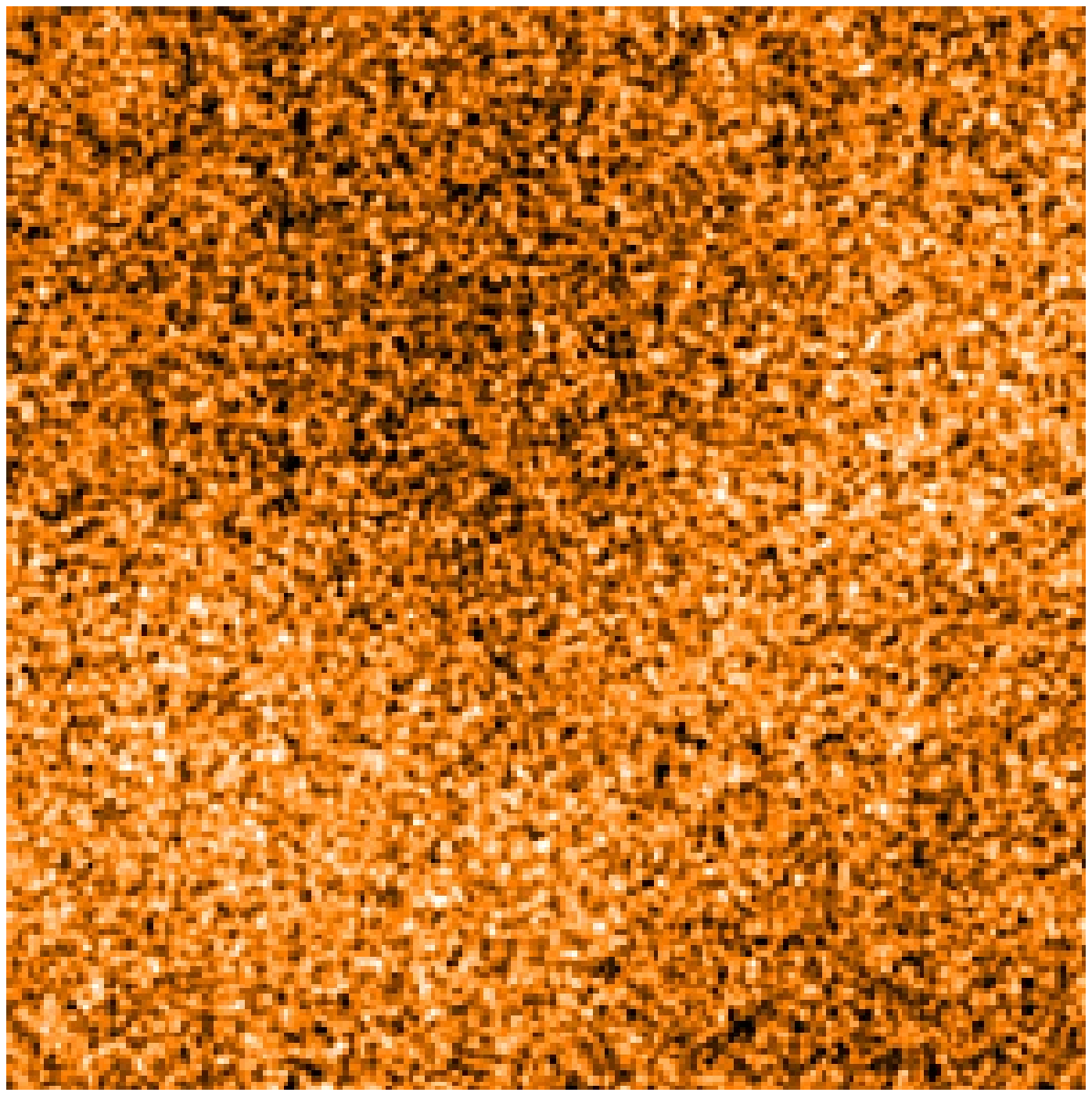}
\includegraphics[width=5.1cm]{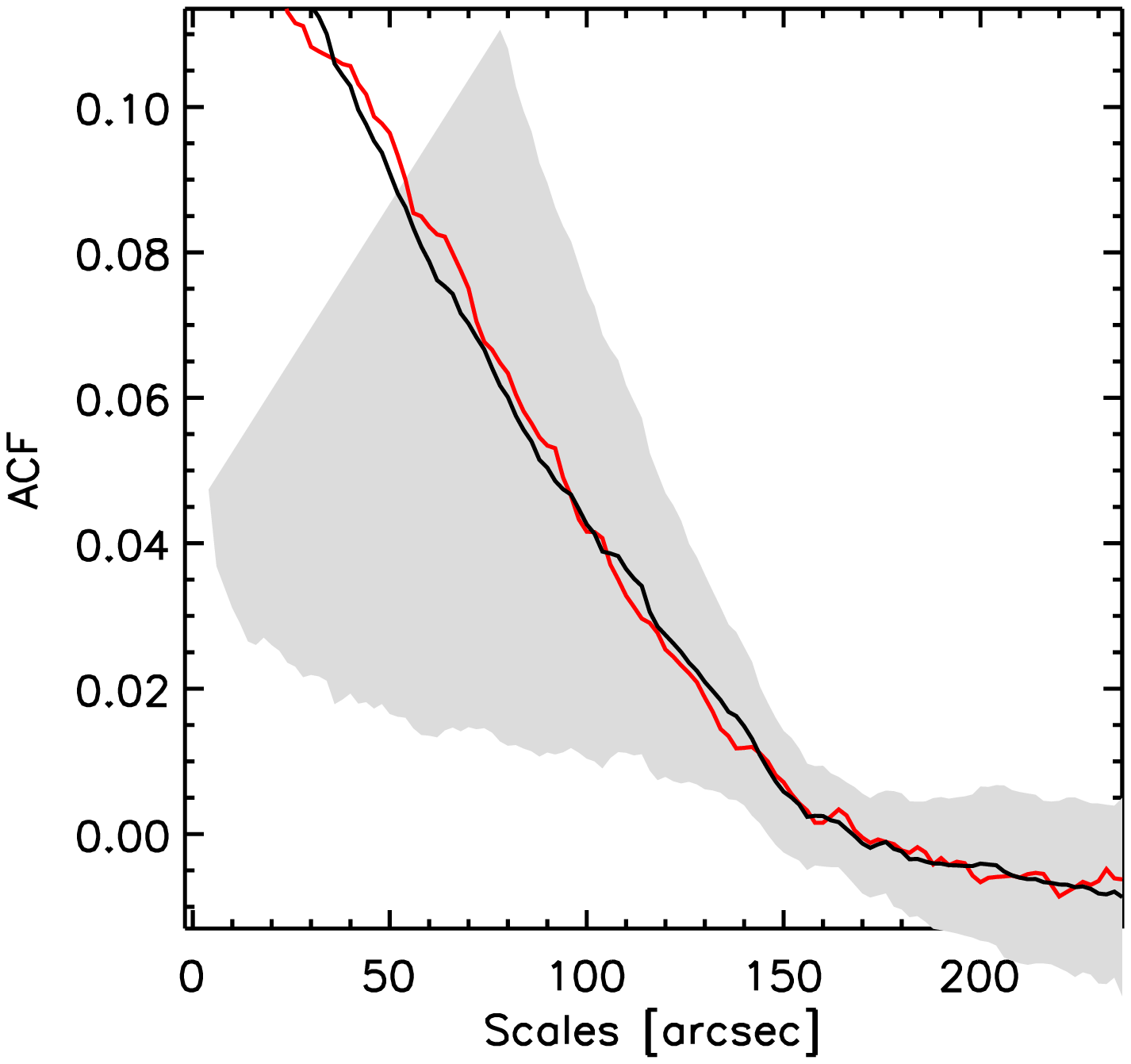}
\includegraphics[width=3.1cm]{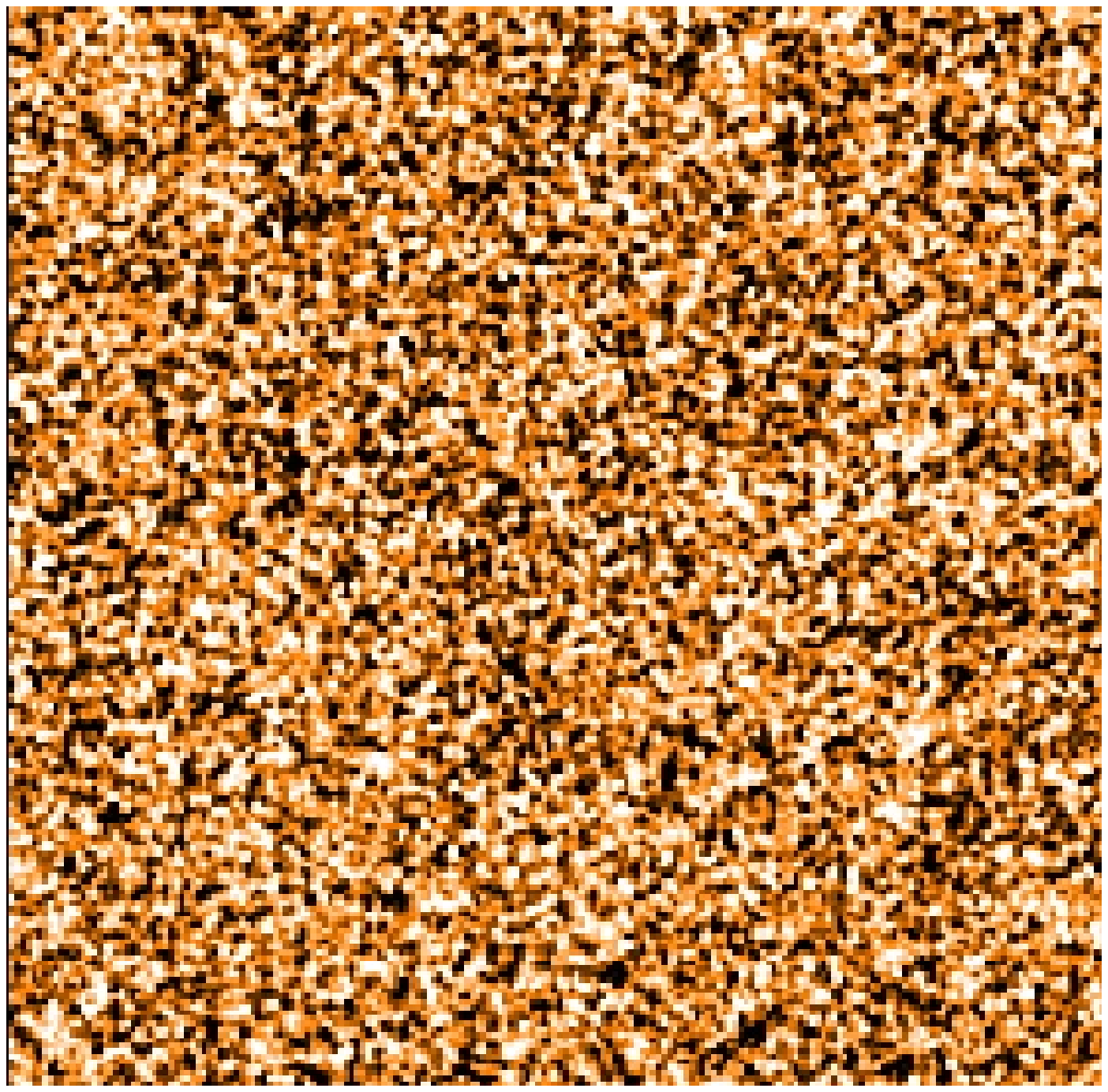}
\includegraphics[width=5.1cm]{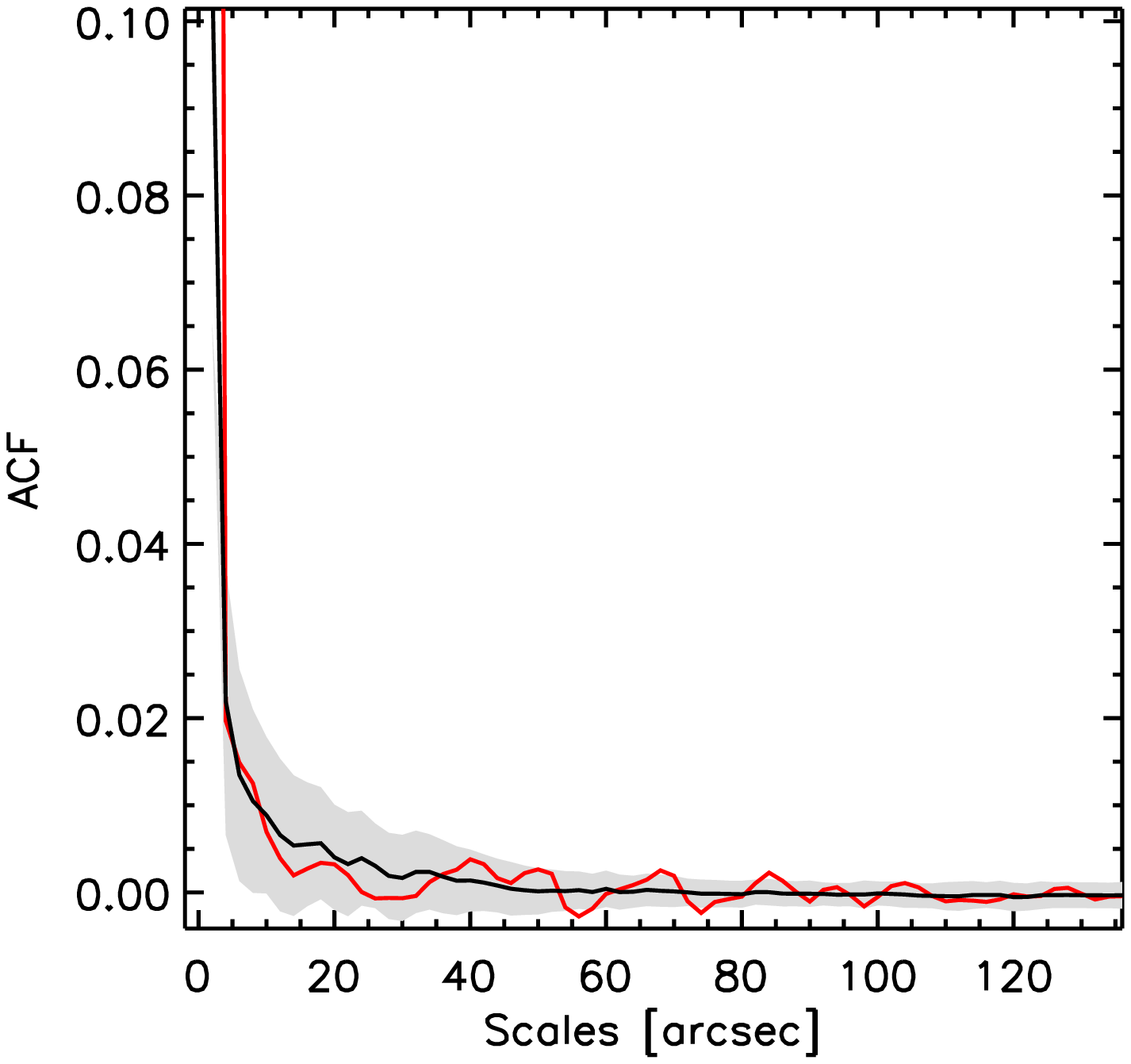}
\end{center}
\caption{\label{FigFakeSky}Two examples of combination of three GRFs
  used to produce fake background fluctuation maps that fit observed
  ACFs. \emph{From left to right:} simulated background map used to
  fit the 2002-10-08 median ACF in Prog. 68.A-0182, corresponding
  median ACF (black curve) with associated scatter (gray area) and
  fitted ACF derived from the simulated map (red line), simulated
  background map used to fit the total 70.A-0591 ACF, corresponding
  ACF with associated uncertainty and fitted ACF. The simulated
  background maps are to be compared visually to those shown in Fig.
  \ref{FigSkyMaps}. Note that the map on the left is one of those that
  show unusually large correlations at all scales (see text).}
\end{figure}

\begin{figure}[h!]
\begin{center}
\includegraphics[height=7.8cm]{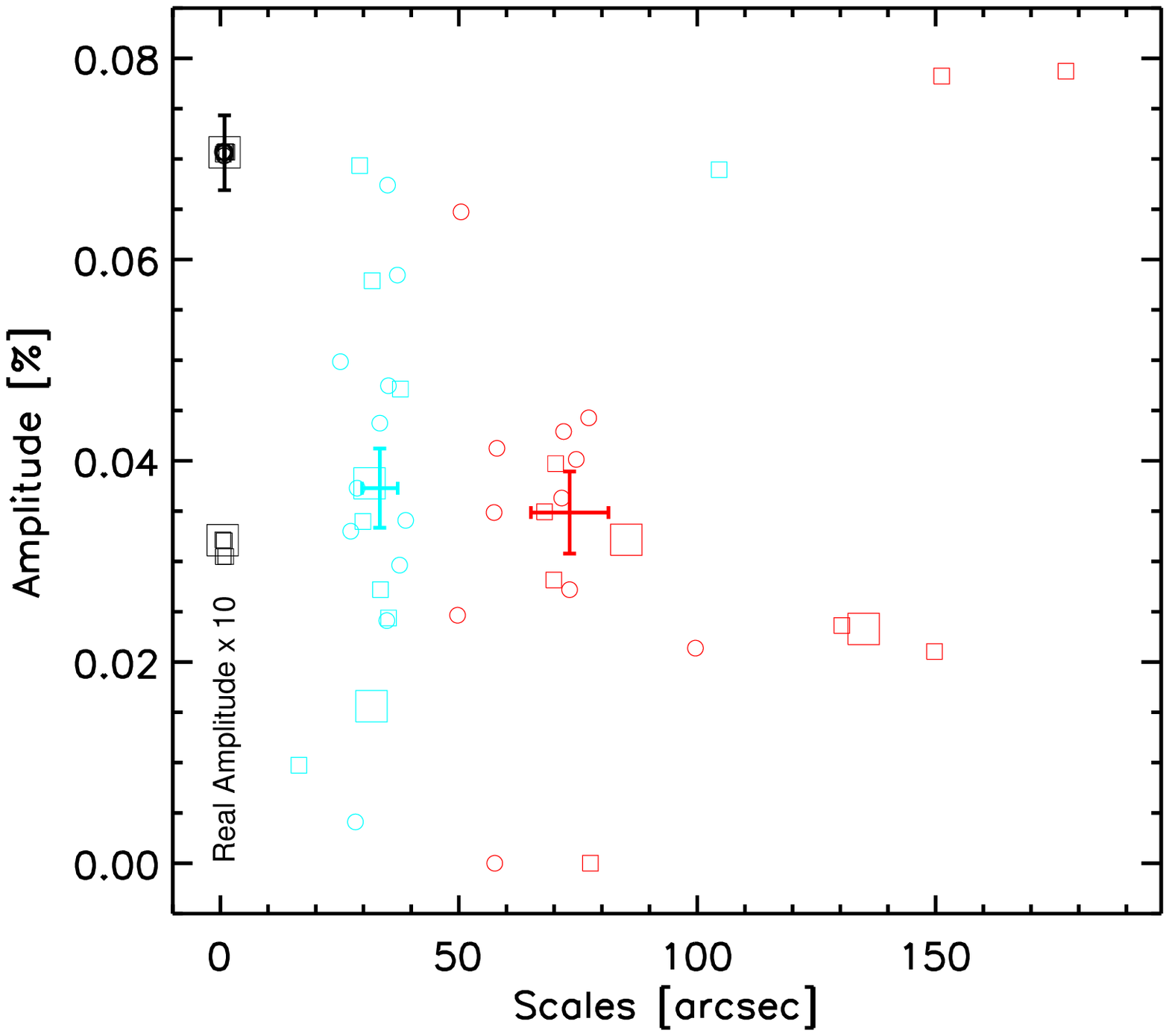}
\includegraphics[height=7.8cm]{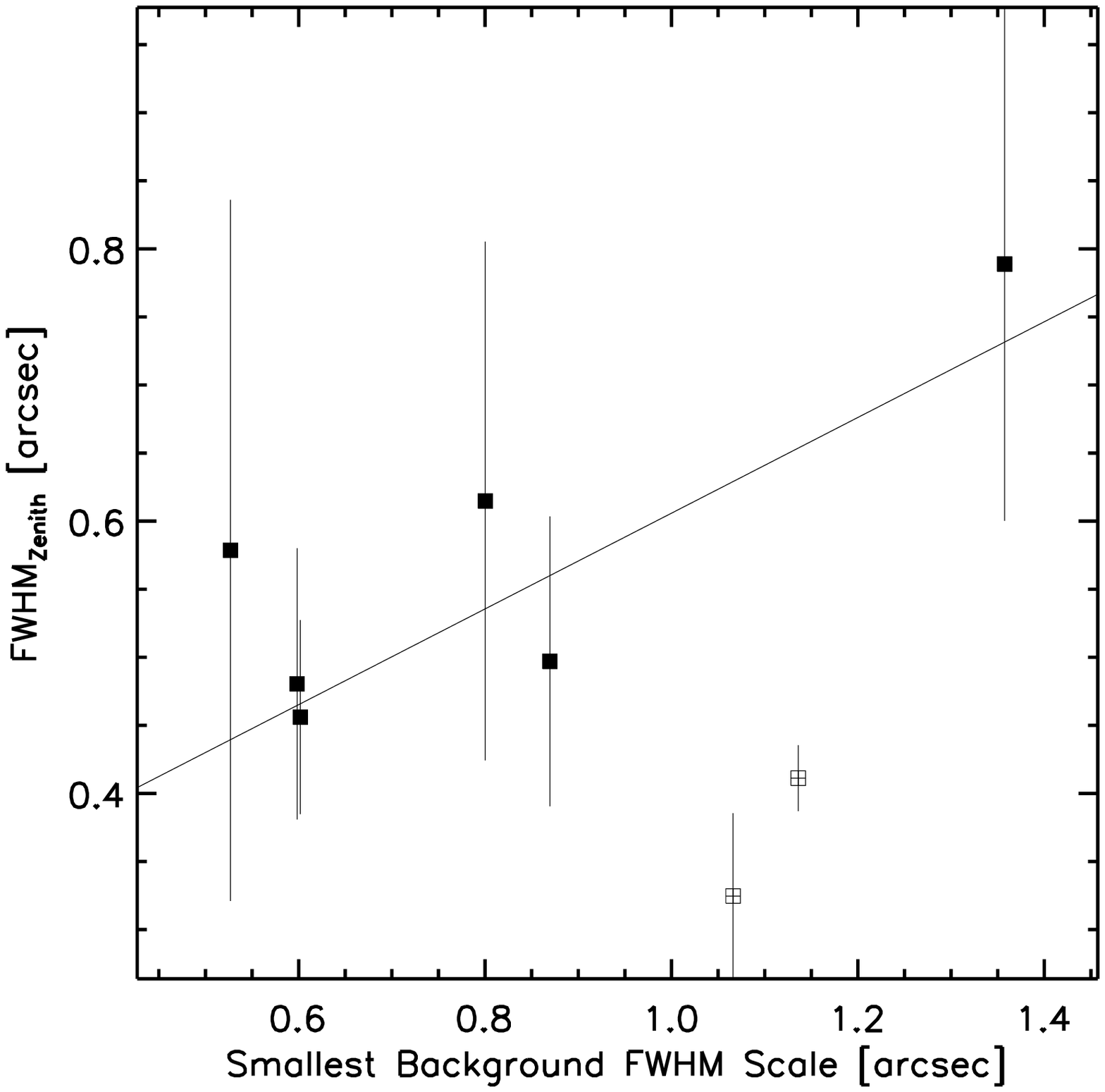}
\end{center}
\caption{\label{FigResSc}\emph{Left:} Result of the fitting of three
  GRFs to the background map ACFs. Amplitudes are given in \% relative
  to the median background value, while scales correspond to the GRF
  FWHMs in arcsec. Black symbols represent the smallest fitted scale
  (their amplitude were divided by a factor ten for visualization
  ease), blue symbols represent the intermediate fitted scale, while
  red symbols represent the largest fitted scale. Empty circles
  represent hourly datasets, small empty boxes represent daily
  datasets, and large empty boxes represent total datasets, while
  crosses represent median values per scale with associated
  uncertainty. \emph{Right:} Expected image quality FWHM at zenith and
  916nm as a function of the smallest FWHM scale (both in arcsec) for
  the daily datasets. The error-bars represent the 1-$\sigma$ scatter
  derived within each dataset, while the black line is a linear fit to
  the data, taking into account only full symbols. }
\end{figure}

The fitted scales and amplitudes of all observed median ACFs are shown
in the left panel of Fig. \ref{FigResSc}. Scales are directly inferred
from the fitted GRF FWHMs, while amplitudes are derived using the
respective weight of each GRF to the total variance of the surface
used to fit the ACF. To translate these relative weights into absolute
contributions in fractions of the median observed background, we
rescaled each weight by the median r.m.s. of all background maps. Fig.
\ref{FigResSc} suggests that three distinct scales are always present
in the data. The smallest scale, but largest in amplitude, appears to
have limited variations in amplitude centered around two typical
values, i.e., $\sim$0.7 and $\sim$0.32\%. This probably reflects the
fact that background maps for Prog. 70.A-0591 still contain
significant photon noise (see Sect. 3.3) and that the real amplitude
of the smallest scale is closer to 0.3\%, or even smaller. We
emphasize that all absolute levels of amplitude has to be taken as
upper limits only, since they were derived from the measured median
r.m.s. fluctuations of the background maps. As discussed in Sect. 3.3,
we indeed cannot exclude that these maps still contain significant
fluctuations associated to photon noise. The two largest scales reveal
amplitudes which are one order of magnitude smaller. The intermediate
scale at $\sim$30 arcsec appears to be relatively well defined in
size, while the largest one appears to be of the order of $\sim$70
arcsec on average, although the data are scattered over a large range.
The median size and amplitude of the three scales are listed in Tab.
\ref{TabRes}.

\begin{table}[h]
  \caption{Median scales (FWHM in arcsec) and amplitudes (in \% relative to the median background value) of the three dominant scales of the relative background fluctuations as inferred from Fig. \ref{FigResSc}. Note that the amplitude of the smallest is probably closer to $\sim$0.32\%, or even smaller (see text).} 
\label{TabRes}
\begin{center}       
\begin{tabular}{cccc}\hline
 & Smallest scale & Intermediate scale & Largest scale\\\hline
Scale & 0.86$\pm$0.06 & 33.43$\pm$3.65 & 73.23$\pm$8.01\\
Amplitude & 0.71$\pm$0.04 & 0.04$\pm$0.01 & 0.04$\pm$0.01\\\hline
\end{tabular}
\end{center}
\end{table}

\subsection{Possible origin of the fluctuation scales}
To investigate the origin of the smallest scale, we plot in the right
panel of Fig. \ref{FigResSc} the size of the smallest scale as a
function of the mean expected image quality in the corresponding
dataset. Image quality was estimated by converting the seeing values
recorded into the file headers into image FWHM\cite{toko02}. All FWHMs
were derived at $\lambda$=916nm, which is approximatively the center
of the two narrow-band filters considered, and homogenized at zenith
using the mean airmass in the dataset. A turbulence outer scale of 8m
(i.e., the telescope pupil diameter) was used during the
fit\cite{toko02}. If one discards the two points with smallest image
quality FWHM, the probability that the two quantities are correlated
is found to be $\sim$92\%. This suggests that the smallest scale of
the background fluctuations is probably linked to atmospheric
turbulence effects\footnote{which might be related to scintillation.
  Such a possible link is beyond the scope of the present study and
  will be explored in future studies.}.

The largest scale of fluctuation is found to be $\sim$70 arcsec on
average with large scatter. The largest fluctuations (in size) of the
atmosphere are known to be due to gravitational waves in the
mesosphere\cite{high10}. However, such waves reveal wavelengths of
several degrees and r.m.s. variations of $\sim$3-4\%\cite{high10}. It
is therefore unlikely that such fluctuations might be linked to the
largest fluctuations at $\sim$70 arcsec. A more likely explanation for
these fluctuations is a known multiplicative feature that can appear
in the FF frames and varies as a function of the angle of the field
rotator\cite{moehler10}. The number of twilight FF frames available
(see above) is clearly too low to smooth out this pattern, which might
therefore let an imprint in the background maps. Such a pattern is
indeed found to dominate in amplitude the FF variations at scales
larger than $\sim$70 arcsec\cite{moehler10}. This might be a natural
explanation for the large range of values found at the largest scales
($\sim$50 to 180 arcsec), as a result of the random combination of the
field rotator angle between the individual frames and twilight FF
frames.

The intermediate scale is found to be close to $\sim$30 arcsec. Since
we did not correct the fringe pattern explicitly but only applied a
multiplicative correction (whereas the fringe pattern is additive),
such a scale could be associated to fringing residual light. However,
the inter-fringe distance is found to be 15-20 arcsec with amplitudes
$\sim$0.1-0.3\%. These sizes and amplitudes are significantly
different that what is found for the intermediate-scale fluctuations
with FWHM sizes $\sim$25-40 arcsec and amplitudes mostly in the range
0.02-0.07\%. We conclude that our multiplicative correction of the
fringe pattern is accurate enough and that the intermediate-scale
fluctuations of the background are probably not related to fringing.
Moreover, fluctuations at these scales are too small to be associated
to the feature introduced by the field rotator angle as discussed
above\cite{moehler10}. So far, it remains unclear to what process(es)
fluctuations at these scales could be associated, and whether these
fluctuations are associated to sky background fluctuations or
instrument residual light.

\subsection{Limitations and future plans}
One should consider the above results as characterizing the
fluctuations of the global background, i.e., potentially including
residual scattered light or flat field effects, rather than pure sky
background. The possible link between the largest scale fluctuations
and the field rotator scattered light indeed suggests that the
background maps could be polluted with residual instrumental effects.
However, we found that amongst the three scales that emerge
systematically from the ACF associated to these fluctuations, at least
one of them could be associated with atmospheric effects (i.e.,
turbulence). This gives us some confidence that the present results
provide at least upper limits on the sky background relative
fluctuations. To confirm these results, it will be very useful to
analyze more narrow-band imaging data obtained with other instruments.
Finding similar scales and amplitudes with other datasets from another
instrument would strongly suggest that the results of the present
study are not limited by instrument residuals but truly characterize
the real sky background fluctuations.

Finally, direct on-sky tests conducted on fiber-fed spectrographs are
on-going\cite{rodrigues12}. Preliminary results using FLAMES-GIRAFFE
at VLT suggest that a sky background measurement conducted $\sim$10-15
arcsec away from the scientific fiber is sufficient to subtract the
sky signal with an accuracy of one percent. According to the results
of the present study, measurements at such a scale should indeed
provide a signal correlated with all large-scale fluctuations. Sky
subtraction should therefore be limited by fluctuations at the
smallest scale, which will remain hardly measurable in parallel to
scientific observations, even with slit spectroscopy. We found typical
amplitudes of a few tenth of percent at this scale, consistent with a
limit on the sky background subtraction accuracy of $\sim$1\% on the
continuum as suggested by on-going on-sky tests\cite{rodrigues12}. We
emphasize that such on-sky tests were conducted using a fiber-fed
spectrographs, which therefore suggests that such instruments can be
as accurate as slit spectrographs for sky subtraction procedures.

\section{Conclusion}
We have analyzed ESO FORS2 archive narrow-band imaging data at
$\lambda\sim$0.9$\mu m$ to characterize the temporal and spatial
fluctuations of the sky background. We carefully reduced two archive
programmes that sample timescales in the range 0.5-3.5 hr and spatial
scales up to 240 arcsec. The limited signal-to-noise ratio and
possible non-identified residual flat field and/or scattered light
effects are clear limitations to such an exercise. Assuming that upper
limits on the total background fluctuations can nevertheless be
derived, we summarize our results as follows:
\begin{enumerate}
\item Background fluctuations show, as expected, random spatial and
  temporal behaviors. Large variations in the spatial correlation are
  observed at all the sampled timescales. Nevertheless, a dominant
  behavior in the spatial fluctuations appears to emerge.

\item Relative spatial fluctuations of the background (relative to the
  median background value) is found to be dominated by three distinct
  scales. All these fluctuations result in otal r.m.s. variations
  below one percent.

\item The smallest scale dominates in term of amplitude ($\sim$0.3\%),
  which is found to be one order of magnitude larger than the two
  largest scales. This scale could be related to atmospheric
  turbulence.

\item The largest scale fluctuates at amplitudes of $\sim$0.035\%.
  This spatial scale shows large variations in size, with values
  between 50-180 arcsec in the present study. These fluctuations could
  be linked with scattered light depending on the field rotator angle.

\item The intermediate scale is found to be well defined at $\sim$30
  arcsec with amplitudes similar to those of the largest scale. It is
  unlikely that these fluctations are related to fringing or the field
  rotator effect. It remains unclear to what process(es) such fluctuations
  could be associated.

\item From an instrument design point of view, this study suggests
  that sky background continuum fluctuations should be sampled over
  timescales significantly below 30 min and on spatial scales
  significantly below 30 arcsec. Fluctuations at the smallest scale
  are probably limiting any sky background measurement/subtraction
  technique, even with slit spectrographs. We found that fluctuations
  at this scale have amplitudes of a few tenth of percent, which
  likely represent a lower bound to the accuracy one can reach in term
  of sky subtraction with spectroscopic observations.

\item Results from on-sky tests suggest that a limit of 1\% in
  accuracy of the sky subtraction process can already be reached on
  existing red-optical spectrograph, including fiber-fed instruments.
  Further tests will be needed before generalizing this result in the
  NIR regime, in which residual light from thermal emission background
  will be an additional source of additive residual light and will
  therefore further limit the accuracy on can reach of the sky
  subtraction process.
\end{enumerate}


\acknowledgments 
We thank all ESO staff contributing to maintain the raw data archive,
which was extensively used in this study.


\bibliography{report}   
\bibliographystyle{spiebib}   

\end{document}